# Femtosecond control of electric currents at the interfaces of metallic ferromagnetic heterostructures

T. J. Huisman[1], R. V. Mikhaylovskiy[1], J. D. Costa[2,3], F. Freimuth[4], E. Paz[2], J. Ventura[3], P. P. Freitas[2], S. Blügel[4], Y. Mokrousov[4], Th. Rasing[1] and A. V. Kimel[1]

[1]*Radboud University, Institute for Molecules and Materials, 6525 AJ Nijmegen, The Netherlands.*
[2]*International Iberian Nanotechnology Laboratory, INL, Braga, Portugal.*
[3]*IN-IFIMUP, Rua do Campo Alegre, 687, 4169-007 Porto, Portugal.*
[4]*Peter Grünberg Institut and Institute for Advanced Simulation, Forschungszentrum Jülich and JARA, 52425 Jülich, Germany.*

**The idea to utilize not only the charge but also the spin of electrons in the operation of electronic devices has led to the development of spintronics, causing a revolution in how information is stored and processed. A novel advancement would be to develop ultrafast spintronics using femtosecond laser pulses. Employing terahertz ($10^{12}$ Hz) emission spectroscopy, we demonstrate optical generation of spin-polarized electric currents at the interfaces of metallic ferromagnetic heterostructures at the femtosecond timescale. The direction of the photocurrent is controlled by the helicity of the circularly polarized light. These results open up new opportunities for realizing spintronics in the unprecedented terahertz regime and provide new insights in all-optical control of magnetism.**

Despite the fact that the generation of spin-polarized currents with the help of circularly polarized light has been demonstrated in noncentrosymmetric semiconductors [1], an application of this concept to metallic materials which are commonly used in spintronic devices has not yet been realized. The helicity dependent photocurrents can be induced due to the Rashba effect [1], which is one of the microscopic mechanisms representing the coupling between the momentum and spin of an electron [2]. This spin-orbit interaction of the conduction electrons in a system characterized by the lack of space inversion symmetry leads to a breaking of the degeneracy between spin-up and spin-down electron sub-bands in momentum space [3-7]. The Rashba effect can be responsible for the phenomenon of spin-orbit torque, when a spin-polarized electric current flowing through a magnetic conductor induces a torque acting on the magnetization $\mathbf{M}$ and tilts it [8-11]. The phenomenon when a tilting magnetization produces an electric current can be seen as an inverse effect. On the other hand, it has been shown that due to the non-dissipative inverse Faraday effect [12,13] or due to the dissipative optical spin transfer torque effect (OSTT) [14], circularly polarized light can induce a tilt to the magnetization. The direction in which the magnetization is tilted is given by $[\mathbf{M} \times \boldsymbol{\sigma}]$, where $\boldsymbol{\sigma}$ is the axial unit vector pointing parallel or antiparallel to the propagation of light depending on its helicity. Hence if a femtosecond circularly polarized optical excitation acts on the magnetization as an effective magnetic field inducing a torque, it should also produce a spin-polarized current mediated by the Rashba effect. The direction of this current is controlled by the helicity of the employed light. However, despite the fact that various helicity dependent ultrafast effects of light on magnetization have been reported for magnetic dielectrics, semiconductors and amorphous metallic alloys, the feasibility of similar effects in pure metals is still unclear. As a result, the significance of helicity-dependent photocurrents in metallic materials is a priori not obvious.

Although transition ferromagnetic metals do possess space inversion symmetry, effective symmetry breaking can be present at the interfaces between different metals. Interface sensitive magneto-optics [15,16] and the formation of chiral spin structures at the surfaces and interfaces [17-19] are typical examples of magnetic phenomena originating from the spontaneous symmetry breaking at the interfaces of otherwise centrosymmetric metals. For an inverse spin-orbit torque mediated by a magnetization change, the resulting current is perpendicular to both the light-induced magnetization and the direction in which the space inversion symmetry is broken $\mathbf{n}$. In particular, in metallic heterostructures with in-plane structural isotropy and in-plane magnetization, one can anticipate the existence of an interfacial in-plane photocurrent generated by circular polarized light as

$$\mathbf{j} = \chi \mathbf{n} \times [\mathbf{M} \times \boldsymbol{\sigma}] I \qquad (1)$$

where $\chi$ is a scalar, $\mathbf{n}$ is a polar unit vector normal to the interface and $I$ is the intensity envelope of a circularly-polarized light pulse which exerts a torque on the magnetization and thereby tilts it in plane, see Fig. 1a. A similar equation for the helicity dependent photocurrents can be derived phenomenological by assuming that the symmetry of our heterostructure is $C_{\infty v}$ [20]. This photo-induced current does not rely on laser-induced heating, contrary to all previous demonstrations of laser-induced spin-currents in metallic heterostructures [21-25].

For our experiments we studied heterostructures made of a single metallic ferromagnetic (FM) layer and a single metallic non-magnetic (NM) layer deposited on a 0.5 mm thick glass substrate. The structures are very similar to those commonly studied in spintronics related research [4,11,24,26-29]. The FM layer chosen here is a 10 nm thick Co film. The main set of the measurements were performed for the heterostructure with a 2 nm thick Pt NM layer. Supporting experiments were performed using heterostructures in which the NM layer was a 2 nm thick Ru, Ta, and Au layer. We also fabricated a 12 nm thick Co sample without NM layer, which acted as a reference.

The geometry of the experiment is shown in Fig. 1a. To demonstrate the generation and control of the photocurrents, we employ femtosecond circularly polarized laser pulses with a pulse width of 50 fs and a central wavelength of 800 nm. Note that all our experiments are performed at room temperature. According to the Maxwell equations, any sub-picosecond current pulse in the plane of the heterostructure should act as an emitter of electromagnetic radiation in the THz spectral range, polarized parallel to the direction of the current [24], as we also derived in the supplementary information section 1. We detect the electric field of this THz radiation as a probe for the interfacial currents. Further details of the experimental procedure can be found in the Methods section. The set of unit vectors **x**, **y** and **z** represent the chosen coordinate system. The incident laser pulses propagate along **z**, the magnetization is parallel to **y** and the *x* and *y* components of the electric field of the emitted THz radiation are detected.

Note that femtosecond laser-induced emission of the THz radiation polarized perpendicularly with respect to the magnetization, i.e. along **x**, has been reported for similar samples before [24]. The origin of this emission was assigned to electric currents emerging due to the inverse Spin-Hall effect acting on the spin current launched as a result of the ultrafast laser-induced heating of the ferromagnetic metal. As the direction of this spin current is defined by the sample structure and magnetization, no dependence of the corresponding THz emission on the pump polarization was reported in [24] or observed by us. From Eq. (1), as well from considering the Landau-Lifshitz equation (see section 2 of the supplementary information), it is seen that the helicity dependent laser-induced current is launched parallel to the magnetization $\mathbf{M}$, i.e. along the *y*-axis and does not require laser-induced heating. Hence, aiming to demonstrate the interfacial helicity-dependent femtosecond photocurrents, we address our attention to the laser-induced THz emission polarized along **y**.

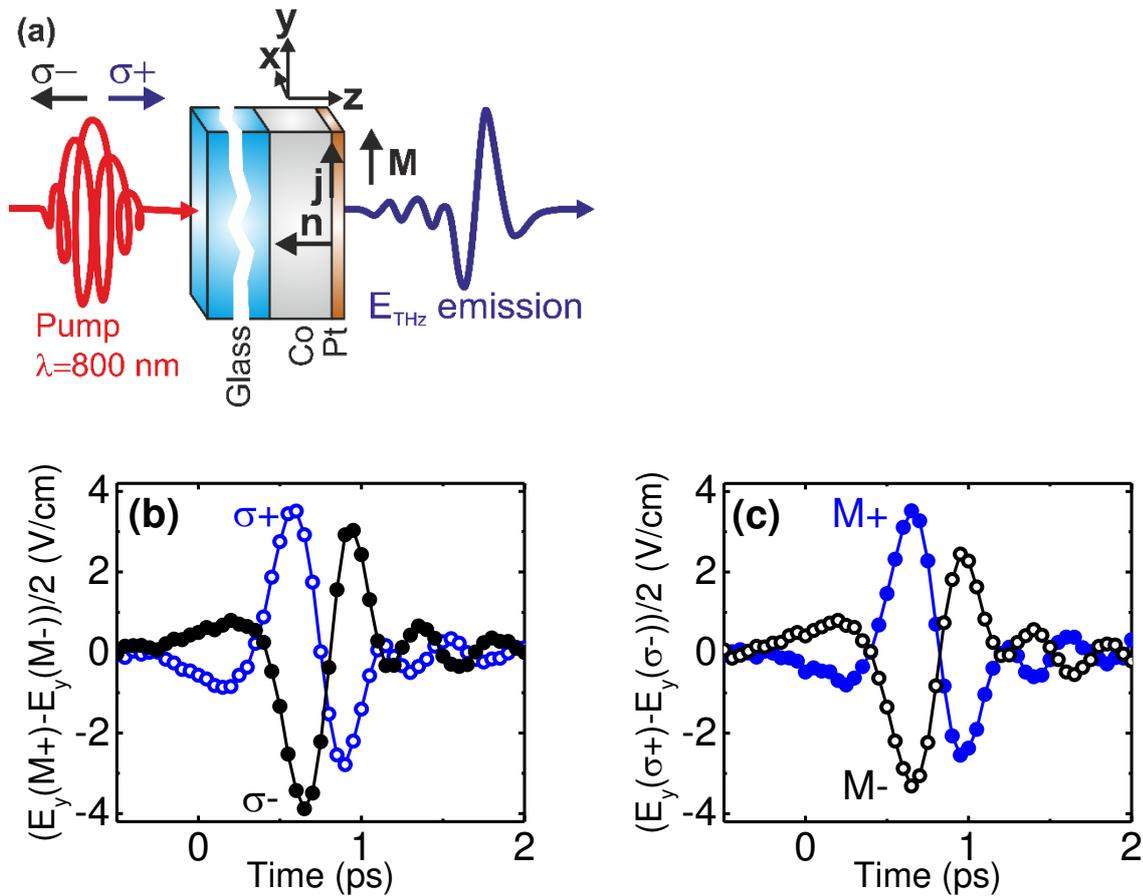

**Figure 1 | Experimental schematics and symmetry of the emitted THz radiation. (a)** Layered structure under study and the scheme of the experiment. A magnetic field B=0.1 Tesla is applied in-plane to saturate the magnetization **M**. **(b)** Electric field of the emitted radiation polarized along the *y*-axis and odd with respect to the magnetization **M** as a function of time measured for the opposite helicities of light. **(c)** The emitted radiation changes sign with magnetization. The position of zero time is arbitrarily chosen.

Figure 1b shows time traces of the *y*-component of the pump-induced THz emission odd with respect to the magnetization **M** for opposite helicities of the pump light. The time-traces were obtained by performing the measurements at a magnetic field of B=0.1 Tesla of two opposite polarities and taking the difference. The figure clearly shows that the electric field of the emitted THz radiation changes sign upon reversal of the helicity. In the supplementary material we show that this helicity dependence exists also for the other samples studied. Fig. 1c shows that the emitted electric field also changes sign upon magnetic field reversal. We found that the emission is still present after a reduction of the applied magnetic field to zero, demonstrating a hysteretic behavior.

To reveal the role of the symmetry breaking in the helicity dependent THz generation process, we performed the measurements for two orientations of the sample by rotating the heterostructure around the magnetization over 180° so that the sign of the polar vector **n** is reversed. Figure 2 shows that this rotation leads to a change in the sign of the emitted helicity dependent THz radiation. The apparent change in delay and timescale of the dynamics upon turning the sample stems from the different propagation of THz radiation and pump light at the wavelength of 800 nm through the glass substrate.

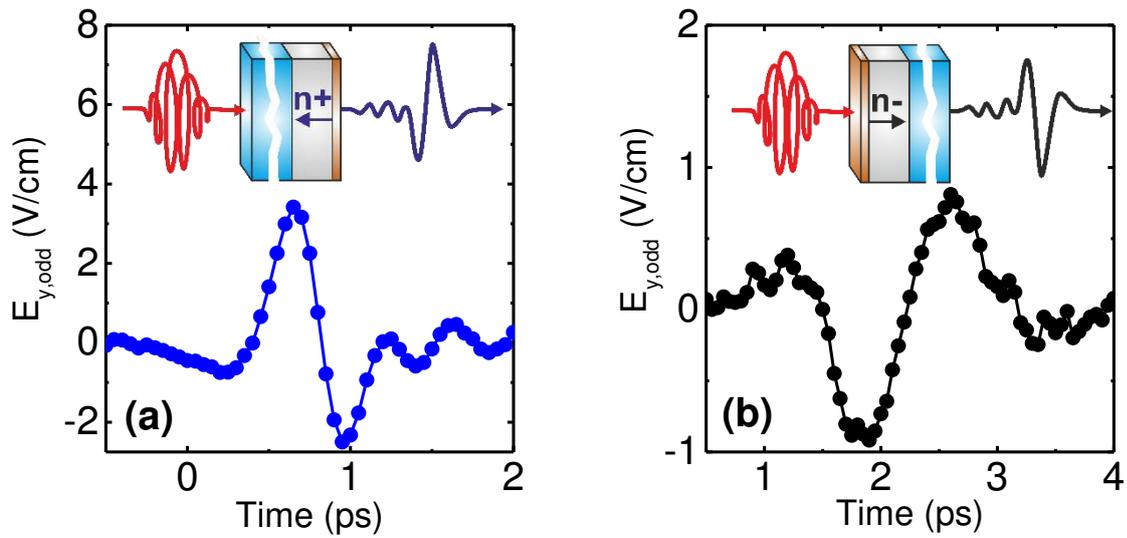

**Figure 2 | The role of the symmetry breaking directionality for the THz emission.** The shown THz emission is odd with respect to the helicity and the magnetization $E_{y,odd}=(E_y(\sigma+,M+)-E_y(\sigma+,M-)-E_y(\sigma-,M+)+E_y(\sigma-,M-))/4$. **(a)** Electric field of the emitted radiation when the pump is incident from the side of the substrate. **(b)** The same as panel (a) but for the case when the pump is incident from the side of the Pt layer. The apparent change in delay and timescale of dynamics when reversing the sample stems from different propagation of THz radiation and light at the wavelength of 800 nm through the substrate. The position of zero time is arbitrary chosen, but kept consistent between the measurements.

Figure 3 demonstrates that the electric field of the THz radiation scales nearly linearly with the intensity and saturates at higher fluences. The linear scaling of the helicity dependent photocurrents reveals that its generation does not rely on laser-induced heating. Unavoidable laser-induced heating results in a conductivity change and a decrease in magnetization. Both of these effects can contribute to the saturation behavior. Our simultaneous measurements of the degree of the photo-induced demagnetization with the help of the magneto-optical Kerr effect show that the magnetization decreases linearly with increasing light intensity (see supplementary information section 4). Finally, the fact that the electrons are excited far above the Fermi level may also affect the efficiency of the spin-orbit-coupling mechanism itself.

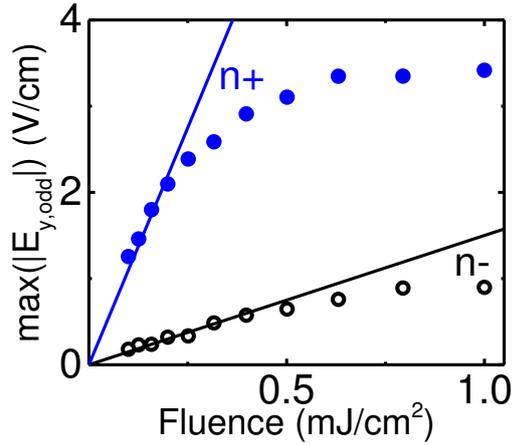

**Figure 3 | Fluence dependence of the peak amplitude of the emitted electric field.** The shown amplitudes are odd with respect to the helicity and the magnetization $E_{y,odd}=(E_y(\sigma+,M+)-E_y(\sigma+,M-)-E_y(\sigma-,M+)+E_y(\sigma-,M-))/4$, and measured for when the pump is incident from the side of the substrate (n+) and Pt layer (n-). The solid lines are linear regressions.

All these experiments indicate that the emitted THz radiation is in full qualitative agreement with the phenomenology of the interfacial helicity-dependent femtosecond photocurrent expressed in Eq. (1). In order to estimate the amplitude of the observed photocurrent, using the Maxwell equations we derived how the photocurrent from Eq. (1) is related to the electric field of the observed terahertz radiation in our spectrometer (see supplementary information section 1). Our estimate gives that the photocurrent rises on a time-scale of approximately 300 fs with a peak amplitude of the order of $10^{11}$ $A/m^2$. Note that in order to come to this estimate, we assumed the current to be localized in a 0.5 nm thick interface region between the two different metal layers and we used permittivity values as reported for bulk metals. The mentioned current density is therefore likely overestimated. Also, the spectral bandwidth of the observed THz pulses are comparable to the one of our spectrometer and therefore the bandwidth of the actual dynamics may be broader. Hence the mentioned rise time of 300 fs should also be regarded as an upper bound and the actual dynamics takes place on a time scale between 50 and 300 fs.

We also studied the laser-induced THz emission from the other heterostructures and the pure Co reference sample. A helicity and magnetization dependent THz emission was detected from all samples, see table 1. The helicity dependence for the Co sample is in agreement with our suggestion that circular light can exert a torque on the magnetization such that $\frac{dM_x}{dt} \neq 0$, resulting in a rapid magnetization change which becomes a source of electromagnetic radiation of magnetic dipole origin. As the complex spectrum of the Co/Pt and Co sample emission match, besides an amplitude factor, we also assume that the torque acting on the magnetization occurs on a time scale of 300 fs. Using the Landau-Lifshitz equation one can estimate that in order to account for the observed THz emission, the strength of a 300 fs pulse representing the effective magnetic field pulse, should have an amplitude of the order of 0.5 Tesla. Previously it was suggested that helicity dependent laser-induced magnetization dynamics could be present in all metallic materials [30-32] but a time-resolved observation of this effect in pure transition metals has not been demonstrated up to now (see supplementary information section 5). Contrary to pure Co much stronger emission was observed from the heterostructures, suggesting that this THz emission must be of electric dipole origin and thus also implying that this emission is due to the interfacial helicity-dependent photocurrents.

|        | $E_y$, odd in helicity of the pump (V/cm) | $E_x$, pump polarization independent (V/cm) |
|--------|---|---|
| Co/Pt  | 3.5 | 53 |
| Co/Au  | 0.27 | 36 |
| Co/Ru  | 1.3 | 15 |
| Co/Ta  | 1.2 | 9.9 |
| Co     | 0.086 | 3.7 |

**Table 1 | Observed maximum amplitudes of emitted electric field.** Values are shown for emission polarized parallel to the *y*-axis (helicity dependent) and parallel to the *x*-axis (polarization independent). The values are for the pump light incident from the side of the substrate.

In table 1 we also show the amplitude of the emitted THz electric fields polarized along the *x*-axis, which are heat driven rather than helicity driven. The data reveals that the emission sources of radiation polarized parallel to the *x* and *y*-axes are not directly correlated. We note that the signs of $E_y$ for the cases of Co/Pt and Co/Ta heterostructures are identical. This is in contrast to the spin-Hall angle, i.e. the coefficient which relates the spin-current to the charge current in the presence of the inverse spin-Hall effect, which has different signs for Pt and Ta [27,33].

Finally we compare the amplitudes of the electric field of the emitted THz radiation with the results of previous studies. For this we developed a theory describing the induced current by an opto-magnetic torque and an inverse spin-orbit coupling. The derivation is shown in section 3 of the supplementary information, which also allowed us to provide an expression for $\chi$ in Eq. (1), see Eq. (S17) in the supplementary information. Using this derivation, we quantify the observed effect in terms of the torkance, which is a parameter characterizing the strength of both direct and inverse spin-orbit torques (see supplementary information section 3). Our derived torkance per unit area are shown in table 2 and compared with similar values known from literature. It is seen that values estimated from our experiments are up to 1 order of magnitude larger. This can be explained by the fact that for these estimates it was assumed that the laser-induced dynamics takes place on a timescale of the above mentioned upper bound of 300 fs. Thereby for these estimates we also used a current density which we also noted to be likely an overestimation.

| Our THz experiment (pC/m) | Other Hz experiments (pC/m) | Theory (pC/m) |
|---|---|---|
| $1 \cdot 10^3$ Co(10)/Pt(2) | 49-77 AlO(2)/Co(0.6)/Pt(3) ref. [11] | 47 O/Co(0.6)/Pt(2.3) ref. [5] <br> 106 Al/Co(0.6)/Pt(2.3) ref. [5] |
| $3 \cdot 10^2$ Co(10)/Au(2) | - | 157 MgO/Fe(0.6)/Au(2) ref. [34] |

**Table 2 | Calculated torkance per unit area for our experiment.** Additionally we provide literature values. The torkance values are accompanied with a description of the analyzed structure, in which the numbers between brackets indicate the film thickness in nanometers.

The demonstrated ultrafast generation and control of spin-polarized currents in metals opens up intriguing opportunities for fundamental studies of spintronic phenomena at the femtosecond time scale. Moreover, we note that in the most successful models explaining all-optical helicity dependent magnetic switching, the role of the spin-orbit coupling has been ignored [35,36]. As a result, the origin of the recently discovered helicity dependent switching in Co/Pt multilayers has not yet been understood [31]. Our finding reveals that the interfacial spin-orbit interaction plays a decisive role in the helicity dependent laser-induced dynamics.

## Methods

**Materials.** The metallic films were deposited on a 0.5 mm thick glass substrate using an ultrahigh vacuum multitarget sputtering system, with a base pressure of $5 \cdot 10^{-8}$ Torr. For the metallic heterostructures a 10 nm thick ferromagnetic layer of Co was deposited, capped with a 2 nm thick non-magnetic layer of Pt, Au, Ru or Ta. As a reference we also deposited a single layer of 12 nm thick Co on a glass substrate. For each material the deposition conditions, i.e. applied current and Argon gas flux were optimized to achieve the best quality and reproducibility of the films. All films were deposited with a deposition rate smaller than 1 angstrom per second for a good control of the thickness. During deposition the samples are rotated to ensure uniformity of the films.

**Experimental setup.** The experimental setup is comparable to the one we used in [37]. An intense laser pulse with a duration of 50 fs and a fluence of approximately 1 mJ/cm$^2$, the pump, is incident on the metallic films with a spot diameter of 1 mm, inducing rapid changes to the magnetization and to the electronic transport properties. According to the Maxwell equations, these dynamics give rise to emission of radiation in the terahertz (THz) spectral range, which can be detected as a probe of these dynamics. This THz radiation propagates first through two wiregrid polarizers, used to measure its polarization. These polarizers have a transmission higher than 95% in a range of 0 to 2 THz and have an extinction ratio for an electric field at 1 THz of $2 \cdot 10^3$. Subsequent, the emission is collected and refocused using two gold coated parabolic mirrors. The radiation is focused onto a ZnTe crystal which is simultaneously gated by pulses from the laser. The electric field of the THz radiation induces birefringence by means of the Pockels effect inside the ZnTe crystal, causing additional ellipticity of the gating laser pulses. Measuring this ellipticity with a balanced bridge detection scheme provides the amplitude of the electric field of the emitted THz radiation.

The magnetization dynamics can also be observed using the magneto-optical Kerr effect (MOKE) detected with a weak laser pulse, the probe. This probe are pulses from the laser with a fluence ten times smaller than that of the pump. While the pump is incident normal to the sample, the probe is incident under an angle of 25°, allowing for spatial separation of the reflected beams. The initially linear polarized probe is measured after reflection on the sample polarization resolved with a balanced bridge detection scheme.

**Acknowldegments**
We would like to thank T. Toonen and A. van Etteger for technical support. We would like to thank A. Brataas, A. Kirilyuk, A. K. Zvezdin and V. V. Bel'kov for fruitful discussions. This work was supported by the Foundation for Fundamental Research on Matter (FOM), the European Unions Seventh Framework Program (FP7/2007-2013) grant No. 280555 (Go-Fast) and No. 281043 (FemtoSpin), Projects No. Norte-070124-FEDER-000070 and FEDER-POCTI/0155, European Research Council grant No. 257280 (Femtomagnetism) and grant No. 339813 (Exchange), and the program "Leading Scientist" of the Russian Ministry of Education and Science (14.Z50.31.0034). J. D. C. is thankful for FCT grant SFRH/BD/7939/2011.

Supplementary Information

**Femtosecond control of electric currents at the interfaces of metallic ferromagnetic heterostructures**

T. J. Huisman[1], R. V. Mikhaylovskiy[1], J. D. Costa[2,3], F. Freimuth[4], E. Paz[2], J. Ventura[3], P. P. Freitas[2], S. Blügel[4], Y. Mokrousov[4], Th. Rasing[1] and A. V. Kimel[1]

[1]*Radboud University, Institute for Molecules and Materials, 6525 AJ Nijmegen, The Netherlands.*
[2]*International Iberian Nanotechnology Laboratory, INL, Braga, Portugal.*
[3]*IN-IFIMUP, Rua do Campo Alegre, 687, 4169-007 Porto, Portugal.*
[4]*Peter Grünberg Institut and Institute for Advanced Simulation, Forschungszentrum Jülich and JARA, 52425 Jülich, Germany.*

### I. THz emission derived from the Maxwell equations

Here we derive mathematical expressions describing how a fast laser induced current pulse gives rise to the emitted THz radiation. We assume an heterostructure is in the *xy* plane as indicated in Fig. S1. From Ampere's and Faraday's laws we can relate the electric field of the emitted electromagnetic radiation to the current of the medium, which in Gaussian units and in frequency domain can be expressed as:

$$\frac{\partial^2 \tilde{E}_y(z)}{\partial z^2} + \frac{\omega^2}{c^2}\varepsilon(z)\tilde{E}_y(z) = -\frac{4\pi i \omega}{c^2}\tilde{J}_y(z),  \qquad (S1)$$

where $\tilde{E}_y$ is the *y*-component of the electric field, $\varepsilon(z)$ is the dielectric permittivity, $\omega$ is the angular frequency, $c$ is the speed of light in vacuum, $\tilde{J}_y$ is the *y*-component of the current density and the ~ symbol is used to indicate the Fourier transform with respect to time.

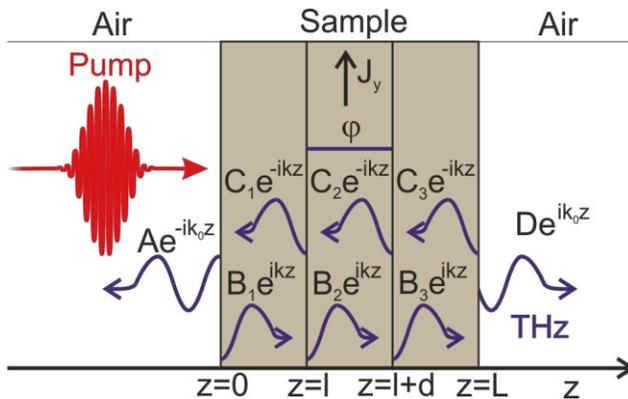

**Fig. S1 | Geometry of the modelled experiment.** A laser pump pulse initiates current dynamics at the interface of two metallic layers, $l < z < (l+d)$. The current dynamics in the sample is accompanied by the emission of electromagnetic radiation at terahertz frequencies which is detected.

We consider the structure shown in Fig. S1; three metallic layers with approximately the same refractive index of which the middle layer represents the interface exhibiting current dynamics and the two outer metallic layers are adjacent to vacuum or air. We write the current density as $\tilde{J}_y(z) = \tilde{J}_y[\Theta(z-l) - \Theta(z-l-d)]$ where $\Theta$ is the Heaviside function, $l$ is the thickness of the most left metallic layer and $d$ is the thickness of the layer in which the current is localized. As shown in Fig. S1, we are looking for the solutions in the form of the sum of the fundamental solution of the homogeneous equation and a partial solution of inhomogeneous equation (S1), from which follows:

$$\tilde{E}_y(z) = \begin{cases} A e^{-ik_0 z}, & z < 0 \\ B_1 e^{-ikz} + C_1 e^{ikz}, & 0 < z < l \\ B_2 e^{-ikz} + C_2 e^{ikz} + \varphi, & l < z < (l+d), \\ B_3 e^{-ikz} + C_3 e^{ikz}, & (l+d) < z < L \\ D e^{ik_0 z}, & z > L \end{cases} \quad (S2)$$

where $\varphi$ is a partial solution, $k_0$ is the wavevector in air, $k$ is the wavevector in the metallic layers and $L$ is the thickness of the whole heterostructure. Taking the partial solution as a constant, its solution is found from eq. (S1) as $\varphi = -\dfrac{4\pi i}{\omega \varepsilon_{metal}} \tilde{J}_y \approx -\sigma^{-1} \tilde{J}_y$, with $\varepsilon_{metal}$ the permittivity of the layer where the current is situated and $\sigma$ the corresponding conductivity, in accordance with Ohm's law. From Faraday's law we know that the electric field of eq. (S2) at the interfaces should be continuous. Furthermore, from eq. (S1) follows that in the absence of delta functions, the derivative of the electric field at the interfaces should also be continuous. This provides the boundary conditions:

$$\tilde{E}_y(z) = \begin{cases} A = B_1 + C_1 \\ -k_0 A = -k B_1 + k C_1 \\ B_1 e^{-ikl} + C_1 e^{ikl} = B_2 e^{-ikl} + C_2 e^{ikl} + \varphi \\ -B_1 e^{-ikl} + C_1 e^{ikl} = -B_2 e^{-ikl} + C_2 e^{ikl} \\ B_2 e^{-ik(l+d)} + C_2 e^{ik(l+d)} + \varphi = B_3 e^{-ik(l+d)} + C_3 e^{ik(l+d)} \\ -B_2 e^{-ik(l+d)} + C_2 e^{ik(l+d)} = -B_3 e^{-ik(l+d)} + C_3 e^{ik(l+d)} \\ B_3 e^{-ikL} + C_3 e^{ikL} = D e^{ik_0 L} \\ -k B_3 e^{-ikL} + k C_3 e^{ikL} = k_0 D e^{ik_0 L} \end{cases} \quad (S3)$$

Solving these equations for $D$, i.e. the emitted electric field amplitude, provides us with

$$D = -\sigma^{-1} J_y k \frac{-e^{ikl}(k+k_0) + e^{-ikl}(k-k_0) + e^{ik(l+d)}(k+k_0) - e^{-ik(l+d)}(k-k_0)}{e^{-ikL}(k-k_0)^2 - e^{ikL}(k+k_0)^2}.$$

(S4)

Since we are working with thin layers, we can approximate (S4) using a Taylor expansion of the form $e^x \approx 1 + x$, which gives us

$$D \approx -\sigma^{-1} J_y \frac{ik^2 d}{2k_0 + iL(k^2 + k_0^2)}. \tag{S5}$$

As the metals we are working with have imaginary permittivity's in the order of $10^5$ around 1 THz [1], while $L$ is in the order of $10^{-8}$ m, we recognize that the emitted radiation can be further approximated as

$$D \approx -\sigma^{-1} J_y \frac{d}{L}, \tag{S6}$$

i.e., our emitted radiation is approximately given by Ohm's law in which the current density is renormalized to an average current density over the whole thickness of the metallic structure. In order to convert the emitted spectrum into the detected spectrum, the spectrometer response needs to be taken into account. In [2] we describe this spectrometer response for our setup in detail.

To estimate the strength of the photocurrent we assume that the current arising from the inversion symmetry breaking occurs only in an interface layer with a thickness of the order of 0.5 nm, corresponding to $d = 0.5$ nm. For the case of magnetic centrosymmetric metals, an interface layer with such a thickness was reported to be the source of the magnetization induced second harmonic generation, a phenomenon which also requires inversion symmetry breaking [3]. Further we took $l = 10$ nm and $L = 12$ nm. As our spectrometer bandwidth is between 0.1 and 3 THz, we are not able to fully reconstruct $J_y(t)$. Rather, we approximate $J_y(t)$ as a Gaussian function with variable width and amplitude. The width of this Gaussian function determines the spectral bandwidth observed and represents the timescale in which the dynamics occur, while the amplitude of this function determines the observed electric field amplitude and represents the maximum current density amplitude. For the Pt and Au heterostructure we approximated $\varepsilon_{metal}$ as $0.5(\varepsilon_{Co} + \varepsilon_{Pt})$ and $0.5(\varepsilon_{Co} + \varepsilon_{Au})$, respectively, where the permittivity values are taken from [1]. Using the exact Maxwell solution in Eq. (S4), we arrive to an estimate for the helicity dependent current pulse in the Co/Pt and Co/Au heterostructures. The maximum current density amplitude is $1.7 \cdot 10^{11}$ and $5 \cdot 10^{10}$ A/m$^2$ for the Co/Pt and Co/Au heterostructures, respectively. For both calculations a full-width-half-maximum of the current pulse of about 330 fs was used.

## II. Helicity dependent magnetization dynamics

Here we show using the Landau-Lifshitz equation that an effective field $B_{eff}$ induced by circular polarized light acting on the magnetization, will only result in significant helicity dependent magnetization dynamics perpendicular to the original magnetization. We can write the Landau-Lifshitz equation for the in-plane magnetization components as:

$$\frac{dM_y}{dt} = -\gamma M_x B_{eff} \tag{S7}$$

$$\frac{dM_x}{dt} = \gamma M_y B_{eff},$$

with $\gamma$ being the electron gyromagnetic ratio. Before the light pulse arrives at the material, $M_y = M_0$ and $M_x = 0$. The magnetization after the light pulse is then given by:

$$M_y = M_0 \cos(\gamma B_{eff} \tau) \quad \text{(S8)}$$
$$M_x = M_0 \sin(\gamma B_{eff} \tau),$$

where $\tau$ is the time duration and $B_{eff}$ is a characteristic value of the effective field induced by light. The rotation of the spin direction is given by $\theta = \gamma B_{eff} \tau$. If $\theta << 1$, we can use a Taylor expansion to write

$$M_y \approx M_0 - \frac{M_0 \theta^2}{2} \quad \text{(S9)}$$
$$M_x \approx M_0 \theta.$$

From this Taylor expansion it is clear that $M_y$ is even with respect to $B_{eff}$, or equivalent even with respect to the helicity of light, while $M_x$ is odd with respect to the helicity of light. The Taylor expansion also shows that for small spin rotations $M_x$ changes faster than $M_y$. So while also helicity dependent emission may be expected for *x*-axis polarized emission, this emission is not odd with respect to the helicity and is smaller compared to the *y*-axis polarized emission.

From the emitted THz radiation from Co we can make an estimate for $\theta$. The $E_x$ amplitude of Co in table 1 of the main text, corresponds purely to a dynamical change of the length of the magnetization vector, i.e. laser induced demagnetization [2]. The $E_y$ amplitude of Co in the same table, corresponds to the simultaneous change of magnetization in the *x* direction, as given by the Landau-Lifshitz equation. The complex spectrum of both $E_x$ and $E_y$ are besides an amplitude factor, identical, hence we can assume they exhibit identical dynamics. From this we obtain a maximum rotation of the magnetization of Co, i.e. $\theta \approx \tan(E_x/E_y) \approx 1.3°$. As the spectra of all samples are comparable, we can also assume that the magnetization dynamics occur on a similar time scale as the current dynamics, i.e. $\tau \approx 300$ fs. From $B_{eff} = \theta/\gamma\tau$ follows an estimate of the $B_{eff}$ close to 0.5 Tesla. Note that as the emission from the heterostructures are not dominated by the magnetization dynamics, but by the current dynamics, we cannot assume that the ratio of the emitted electric field along orthogonal polarization directions to represent the magnetization rotation.

### III. Estimating the torkance of our experiment
When an in-plane electric field **E** is present in a magnetic bilayer structure, such as Co/Pt, the spin-orbit coupling provides a spin-orbit torque **T** which acts on the magnetization. This torque can be expressed as [4]
$$\mathbf{T} = \mathbf{t(M)E}, \quad \text{(S10)}$$

where $\mathbf{t}(\mathbf{M})$ is the torkance tensor which depends on the magnetization $\mathbf{M}$. In a magnetic bilayer structure the torque is approximately determined by two torkance terms which are respectively odd and even functions of magnetization [4,5]:

$$\mathbf{T} = t_{odd}(\mathbf{n}\times\mathbf{E})\times\frac{\mathbf{M}}{M_0} + t_{even}\frac{\mathbf{M}}{M_0}\times\left[(\mathbf{n}\times\mathbf{E})\times\frac{\mathbf{M}}{M_0}\right]. \quad (S11)$$

with $\mathbf{n}$ the normal to the interface and $M_0$ the saturation magnetization. The corresponding torkance tensor can be written as

$$\mathbf{t}(\mathbf{M}) = \sum_i \frac{t_{odd}}{M_0}[(\mathbf{n}\times\mathbf{e}_i)\times\mathbf{M}]\mathbf{e}_i^T + \frac{t_{even}}{M_0^2}\mathbf{M}\times[(\mathbf{n}\times\mathbf{e}_i)\times\mathbf{M}]\mathbf{e}_i^T \quad (S12)$$

where $\mathbf{e}_i$ is a set of mutually orthogonal unit vectors (i=1,2,3) such that $\sum_i \mathbf{e}_i\mathbf{e}_i^T = \mathbf{I}$, with $\mathbf{I}$ being the unit matrix.

Conversely, an electric current density $\mathbf{J}$ is induced when the magnetization direction changes as a function of time. As this phenomenon is the reciprocal of a spin-orbit torque, it can be expressed by the same torkance tensor [6]:

$$\mathbf{J} = \frac{1}{M_0^2 V}\mathbf{t}(-\mathbf{M})^T\left[\mathbf{M}\times\frac{d\mathbf{M}}{dt}\right], \quad (S13)$$

with $V$ the volume of the interfacial current layer, given by $V = Ad$ where $d$ is the thickness of the interfacial layer. The reciprocity between spin-orbit torques and currents induced by magnetization has been demonstrated experimentally in (Ga,Mn)As, for both the odd and even torkance components with respect to magnetization [7]. Inserting Eq. (S12) into (S13) and using $\mathbf{M}\times\frac{d\mathbf{M}}{dt} = -\gamma M_0^2 \mathbf{B}_{eff}$ as follows from Eq. (S8), we obtain for the current density odd with respect to magnetization:

$$\mathbf{J}_{odd} = \frac{\gamma t_{odd}}{M_0 V}\sum_i \mathbf{e}_i[(\mathbf{n}\times\mathbf{e}_i)\times\mathbf{M}]^T \mathbf{B}_{eff}, \quad (S14)$$

or equivalent

$$\mathbf{J}_{odd} = -\frac{\gamma t_{odd}}{M_0 V}\mathbf{n}\times[\mathbf{M}\times\mathbf{B}_{eff}]. \quad (S15)$$

Similarly, one obtains an expression for the current density even with respect to magnetization:

$$\mathbf{J}_{even} = -\frac{\gamma t_{even}}{M_0^2 V}\mathbf{n}\times[\mathbf{M}\times(\mathbf{M}\times\mathbf{B}_{eff})]. \quad (S16)$$

In our experiment $\mathbf{n}$ is along the $z$ axis, $\mathbf{M}$ along the $y$ axis and $\mathbf{B}_{eff}$ along the $z$ axis. Hence, $\mathbf{J}_{odd}$ is along the $y$ axis, while $\mathbf{J}_{even}$ is 0 in our experiment, from which we can identify $\mathbf{J}_{odd} = J_y$. Comparing Eq. (1) in the main text and Eq. (S15) provides us with the relation

$$\chi\sigma I = -\frac{\gamma t_{odd}}{M_0 V}\mathbf{B}_{eff}. \quad (S17)$$

Eq. (S15) can provide us with an estimate for the torkance per interfacial area:

$$\frac{t_{odd}}{A} = -\frac{J_y d}{\gamma B_{eff}} \quad (S18)$$

where we used $V = Ad$. In section 1 of this supplementary, we estimated the current density $J_y$ of the Co/Pt and Co/Au samples to be $1.7 \cdot 10^{11}$ and $5 \cdot 10^{10}$ A/m² respectively, in a layer thickness $d$ of approximately 0.5 nm. From our polarization resolved data on Co in combination with Eq. (S8) assuming $\tau \approx 300$ fs like for our current pulses, we estimate $\gamma B_{eff}$ to be approximately $8 \cdot 10^{10}$ Hz. From this follows a torkance per interfacial area of 1 and 0.3 nC/m for the emission of the Co/Pt and Co/Au samples, respectively.

### IV. Fluence dependence demagnetization

Figure S2 shows the maximum demagnetization as observed with the MOKE probe. This figure demonstrates that the demagnetization scales linearly with the intensity of light for the Co and Co/Pt samples.

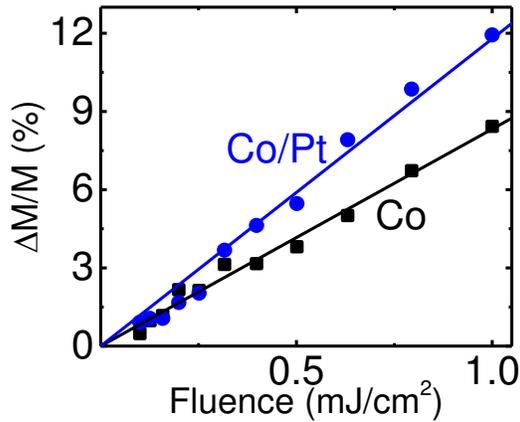

**Fig. S2 | Fluence dependence of the maximum light induced demagnetization.** The measurements were performed using the magneto-optical Kerr effect (MOKE). The pump was incident from the side of the metallic layers deposited on the glass substrate. The solid lines are linear regressions.

### V. Helicity dependence of all samples

Figure S3 shows time traces of the *y*-component of the pump-induced THz emission odd with respect to the helicity and the magnetization. This figure clearly shows that the electric field of the emitted THz radiation of all samples does depend on the helicity of the pump light.

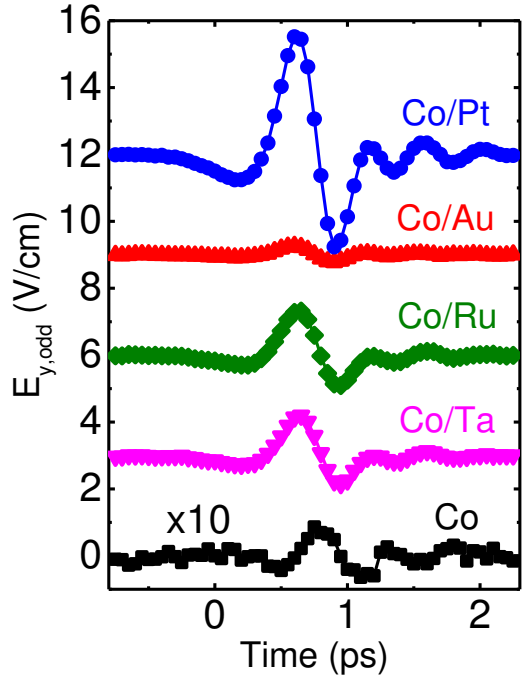

**Fig. S3 | Helicity dependent electric field of the emitted radiation.** The shown emission is polarized along the y-axis, odd with respect to the helicity and the magnetization $E_{y,odd}=(E_y(\sigma+,M+)-E_y(\sigma+,M-)-E_y(\sigma-,M+)+E_y(\sigma-,M-))/4$ and shown for different samples. The amplitude of the emission from the Co sample is multiplied 10 times. The position of zero time is arbitrary chosen and kept consistent relative between the different samples.